# Adaptive Robust Unscented Kalman Filter for Dynamic State Estimation of Power System

Duc Viet Nguyen, Haiquan Zhao, *Senior Member, IEEE*, Jinhui Hu, and Le Ngoc Giang

*Abstract*— Non-Gaussian noise and the uncertainty of noise distribution are the common factors that reduce accuracy in dynamic state estimation of power systems (PS). In addition, the optimal value of the free coefficients in the unscented Kalman filter (UKF) based on information theoretic criteria is also an urgent problem. In this paper, a robust adaptive UKF (AUKF) under generalized minimum mixture error entropy with fiducial points (GMMEEF) over improve Snow Geese algorithm (ISGA) (ISGA-GMMEEF-AUKF) is proposed to overcome the above difficulties. The estimation process of the proposed algorithm is based on several key steps including augmented regression error model (AREM) construction, adaptive state estimation, and free coefficients optimization. Specifically, an AREM consisting of state prediction and measurement errors is established at the first step. Then, GMMEEF-AUKF is developed by solving the optimization problem based on GMMEEF, which uses a generalized Gaussian kernel combined with mixture correntropy to enhance the flexibility further and resolve the data problem with complex attributes and update the noise covariance matrix according to the AREM framework. Finally, the ISGA is designed to automatically calculate the optimal value of coefficients such as the shape coefficients of the kernel in the GMMEEF criterion, the coefficients selection sigma points in unscented transform, and the update coefficient of the noise covariance matrices fit with the PS model. Simulation results on the IEEE 14, 30, and 57-bus test systems in complex scenarios have confirmed that the proposed algorithm outperforms the MEEF-UKF and UKF by an average efficiency of 26% and 65%, respectively.

## I. INTRODUCTION

An accurate dynamic state estimation (DSE) method is essential in the management and monitoring of power systems (PS). To satisfy the increasing requirements, the challenges and research orientations have been mentioned [1]. A summary table of estimated methods has been introduced in [2]. Among them, the method of using the Kalman filter attracts great attention. Additionally, to control noise in the DSE task, random Fourier filter-based filtered-x generalized hyperbolic secant function (RF-FxGHSF) algorithms have proven their effectiveness, especially in handling impulse noise [3].

However, classical Kalman filters (KF), which are developed based on the Gaussian assumption, are confirmed to be ineffective in the face of non-Gaussian noise and outliers. In recent years, Kalman filters based on learning criteria such as maximum correntropy criterion (MCC) and minimum error entropy (MEE) criterion, have achieved superior performance in non-Gaussian noise and outlier environments [4-6]. The reason is that these learning criteria can consider higher-order statistics of the data and robustness to outliers. At the same time, studies have also confirmed that the MEE criterion handles non-Gaussian noise better than the MCC criterion [5]. However, because of the inverse operation of the singular matrix, the MEE criterion suffers from numerical stability problems [7]. To solve this problem, correntropy is added to the error entropy, which is obtained known as the minimum error entropy with fiducial points (MEEF) criterion [8]. Specifically, putting the MCC into the MEE cost function will help MEEF automatically locate the vertex of the error probability density function (PDF) and fix it at the origin.

It can be observed that the above information-theoretic criterion (ITL)-based estimation algorithms all use the traditional Gaussian kernel, which has a fixed shape. Therefore, the resulting error entropy has a fixed shape, making it only able to handle certain types of noise [9]. On the other hand, the generalized Gaussian kernel function is known for its ability to change shape freely. Considering the MCC and MEE criteria, replacing the traditional Gaussian kernels with generalized Gaussian kernels will yield the generalized MCC (GMCC) [10,11,12], and generalized MEE (GMEE) criterion [9,13], respectively, which have better non-Gaussian noise handling capabilities. In addition, some studies have also confirmed that mixture correntropy effectively solves the problem of data with complex distribution properties and further enhances flexibility [14,15]. However, no research has been published to address the problem of inflexibility in the MEEF criterion.

Additionally, an urgent problem for ITL-based estimation algorithms is the value of the kernel shape coefficients. If these values are too large, the performance of these algorithms degrades to just equivalent to the Gaussian assumption. Conversely, if these values are too small, they cannot handle non-Gaussian noise and the fixed-point iteration to find the estimated value will diverge [15]. Furthermore, the scale coefficients that affect the selection of sigma points in unscented transform (UT) also need to be selected [16,17]. It can be seen that the accuracy of the algorithms is directly

This work was partially supported by the National Natural Science Foundation of China (grant: 62171388, 61871461, 61571374). (*Corresponding author*: Haiquan Zhao).

Haiquan Zhao (e-mail: hqzhao_swjtu@126.com), Jinhui Hu (e-mail: jhhu_swjtu@126.com) and Duc Viet Nguyen (e-mail: ruandeyue@my.swjtu.edu.cn), are with School of Electrical Engineering, Southwest Jiaotong University, Chengdu; the Key Laboratory of Magnetic Suspension Technology and Maglev Vehicle, Ministry of Education, China.

Le Ngoc Giang (e-mail: lengocgianglinh@gmail.com ) is with the Faculty of Fundamental Technical, AD-AF Academy, Ha Noi, 12713, Viet Nam.

affected by the values of these coefficients. The selection of these coefficients depends heavily on the number of trials and the experience of the designer. Currently, many studies have discussed and examined the influence of these coefficients on the performance of estimation algorithms [9,13,15-17]. In order to solve the problem of choosing the optimal value of free coefficients, many solutions have been introduced including the application of optimal algorithms [15-17], building a selection process [8], or using adaptive kernels [20]. However, to deal with all the free coefficients, matrices, or finding the missing parameters of the object model simultaneously, meta-heuristic optimization algorithms are always the first choice.

In addition, the uncertainty of the noise distribution also needs to be considered in the dynamic estimation process. The KF has already been explored for dynamic state estimation being a process highly dependent on a prediction step for calculating the system state during the update step. Therefore, the covariance matrices have a great influence on the estimated accuracy [21,22]. Most of the estimates assume that can be calculated accurately in advance the statistics of measurement noise and process noise. In practice, this will have difficulty in dynamic estimates. So, when the covariance matrices are updated at each step time, the estimated accuracy will be ensured. Because measurements at each estimated time will accurately reflect the attributes of the noise. The Sage-Husa estimator is a very effective tool for estimating noise [23,24].

In summary, non-Gaussian noise and outliers; the inflexibility of the traditional Gaussian kernels; the uncertainty of noise distribution; the optimal value of the free coefficients are the main motivations in this paper. Therefore, a robust adaptive UKF (AUKF) under generalized minimum mixture error entropy with fiducial points (GMMEEF) over improve Snow Geese algorithm (ISGA) (ISGA-GMMEEF-AUKF) is proposed to overcome the above difficulties. The main contributions are as follows:

1) The GMMEEF criterion is constructed to overcome the influence of non-Gaussian noise and outliers, which uses a generalized Gaussian kernel combined with mixture correntropy.
2) An AUKF based on GMMEEF criterion (GMMEEF-AUKF) is developed. In which, the uncertainty of noise distribution is considered and the noise covariance matrices are estimated in a numerically stable manner through the modified Sage-Husa estimator.
3) An improve Snow Geese algorithm (ISGA) is designed. Specifically, the exploitation phase is enhanced the accuracy and convergence speed.
4) A robust AUKF under GMMEEF over IGSA (ISGA-GMMEEF-AUKF) is proposed, in which ISGA is utilized to automatically calculate the optimal value of coefficients such as the scale coefficients that affect the selection of sigma points in UT, the kernel shape coefficients, etc.

This paper consists of six main sections: Section II describes the PS model and MEEF criteria; Section III derives the GMMEEF-AUKF algorithm; Section IV derives the ISGA-GMMEEF-AUKF optimal state estimation algorithm; Section V proof of convergence; Section VI reports experimental results; and Section VI concludes.

## II. POWER SYSTEM MODEL AND MEEF CRITERION

In this section, the basic components of the power system and the nature of the MEEF criteria are introduced, to serve the process of building and testing the proposed algorithm in the following sections.

### A. Power System Dynamic Model

This paper employs a PS model assuming that the system operates in a near-steady state [10,15]. Therefore, nonlinear discrete-time equations can be used to illustrate the relationship between measurements and the state variables of the PS model:

$$\begin{cases} \mathbf{u}_t = \mathbf{f}(\mathbf{u}_{t-1}) + \mathbf{q}_t \\ \mathbf{v}_t = \mathbf{g}(\mathbf{u}_t) + \mathbf{r}_t \end{cases} \quad (1)$$

where $\mathbf{u}_t$: the state variable vector dimension $n$ and contains the voltage phase and amplitude of each node at time $t$; $\mathbf{v}_t$ the measurement vector dimension $m$ at time $t$ and contains the magnitude of the voltage of each node; active power injection; active power flow; reactive power injection; reactive power flow; $\mathbf{f}(\mathbf{u}_{t-1})$: state transition function of $\mathbf{u}_{t-1}$; $\mathbf{g}(\mathbf{u}_t)$: measurement function of $\mathbf{u}_t$; $\mathbf{r}_t$ and $\mathbf{q}_t$ represent measurement noise, process noise at time $t$ with covariance matrices $\mathbf{R}_t \in \mathbb{R}^{m \times m}$ and $\mathbf{Q}_t \in \mathbb{R}^{n \times n}$, respectively.

The state transition function $\mathbf{f}(\mathbf{u}_{t-1})$ can be represented using a state prediction method. By employing Holt's two-parameter exponential smoothing technique the function $\mathbf{f}(\mathbf{u}_{t-1})$ is derived as follows:

$$\mathbf{f}(\mathbf{u}_{t-1}) = \mathbf{\Delta}_{t-1} + \mathbf{\Theta}_{t-1} \quad (2)$$

$$\mathbf{\Delta}_{t-1} = \Upsilon \mathbf{u}_{t-1} + (1-\Upsilon) \tilde{\mathbf{u}}_{t-1} \quad (3)$$

$$\mathbf{\Theta}_{t-1} = \Upsilon (\mathbf{\Delta}_{t-1} - \mathbf{\Delta}_{t-2}) + (1-\mho) \mathbf{\Gamma}_{t-2} \quad (4)$$

where $\mho, \Upsilon$ are the coefficients within (0,1); $\mathbf{u}_{t-1}$, $\tilde{\mathbf{u}}_{t-1}$: state vectors, predicted state vector respectively at time $t$-1.

In addition, the measurement function $\mathbf{g}(\mathbf{u}_t)$ represents the real power relationship at time $t$: standard real, power flow equations, and reactive power balance, which is described by the following equations [10,15]:

$$\begin{cases} N_{i,t} = \sum_{j=1}^{N} |V_{i,t}||V_{j,t}|(S_{ij} \cos \varphi_{ij,t} + F_{ij} \sin \varphi_{ij,t}) \\ M_{i,t} = \sum_{j=1}^{N} |V_{i,t}||V_{j,t}|(S_{ij} \sin \varphi_{ij,t} - F_{ij} \cos \varphi_{ij,t}) \\ N_{ij,t} = V_{i,t}^2(S_i + S_{ij}) - |V_{i,t}||V_{j,t}|(S_{ij} \cos \varphi_{ij,t} + F_{ij} \sin \varphi_{ij,t}) \\ M_{ij,t} = -V_{i,t}^2(F_i + F_{ij}) - |V_{i,t}||V_{j,t}|(S_{ij} \sin \varphi_{ij,t} - F_{ij} \cos \varphi_{ij,t}) \end{cases} \quad (5)$$

where $\varphi_{ij,t} = \varphi_{i,t} - \varphi_{j,t}$: the voltage phase difference at time $t$ between the $i$ and $j$ nodes; $F_i$ and $S_i$ represent the susceptance and conductance of the Shunt at node $i$; $F_{ij}$ and $S_{ij}$ represent the susceptance and conductance of the line between the $i$ and $j$ nodes, respectively; $|V|$ denote the voltage magnitude of a node; $M_{i,t}$ and $N_{i,t}$: represent the reactive power and real power at time *t* of node *i*; $M_{ij,t}$ and $N_{ij,t}$: represent the reactive power and real power at time *t* between nodes *i* and *j*.

*B. MEEF Criterion*

In ITL, Renyi's quadratic entropy [5] is defined as follows:
$$H_2(e) = -\log_2(\int (p_e(e))^2 de) \tag{6}$$
where: *e*: error variable; $p_e(.)$: the PDF.

The MEE is proposed by minimizing (6):
$$V(e) = \int (p_e(e))^2 de \tag{7}$$

Utilize the Parzen's estimator, the PDF is given as follows:
$$\hat{p}_e(e) = \frac{1}{N}\sum_{i=1}^{N} G_\sigma(e - e_i) \tag{8}$$
where $G_\sigma(x) = \exp(-x^2/(2\sigma^2))$: Gaussian kernel function with kernel size σ; *N*: the number of errors. Substituting (8) into (7), the MEE obtained by maximizing:
$$\hat{V}(e) = \frac{1}{N^2}\sum_{i=1}^{N}\sum_{j=1}^{N} G_\sigma(e_j - e_i) \tag{9}$$

Since MEE in (9) only minimizes the differences among errors. Error may on the line of 5π/4 and π/4, not be located at zero after optimization. To an automatic way set the error at zero, the MCC is put into MEE [7,8].
$$J = \kappa\sum_{i=1}^{N} G_{\sigma_1}(e_i) + (1-\kappa)\sum_{i=1}^{N}\sum_{j=1}^{N} G_{\sigma_2}(e_j - e_i) \tag{10}$$
with the scaling factor $\kappa \in [0,1]$ and kernel sizes $\sigma_1$ and $\sigma_2$.

*Remark 1*: When $\kappa = 1$, the MEEF criterion is simplified into the MCC criterion [6]; when $\kappa = 0$, the MEEF criterion is simplified into the MEE criterion [5].

## III. GMMEEF-AUKF ALGORITHM

In this section, the limitations of MEEF are analyzed and the uncertainty of noise distribution is considered. A robust optimization criterion is constructed simultaneously an adaptive estimated algorithm is developed.

*A. GMMEEF Criterion*

The Gaussian kernel with its shape cannot be changed freely, which is the reason that makes MEEF less flexible and adaptable. On the other hand, the generalized Gaussian kernel function with its shape can be freely adjusted [9-13,15]. Furthermore, when two generalized Gaussian kernels are fused (mixture correntropy), it will increase flexibility, adaptation and resolve the data problem with complex attributes [14,15]. Combining the above analysis, a new optimal criterion, namely generalized minimum mixture error entropy with fiducial points (GMMEEF) is constructed. The cost function of GMMEEF is defined as follows:
$$J = \kappa\sum_{i=1}^{N}\left[\phi G_{\alpha_1,\beta_1}(e_i) + (1-\phi)G_{\alpha_2,\beta_2}(e_i)\right] + \\ + (1-\kappa)\sum_{i=1}^{N}\sum_{j=1}^{N} G_{\alpha_3,\beta_3}(e_j - e_i) \tag{11}$$

$$G_{\alpha_i,\beta_i} = \frac{\alpha_i}{2\beta_i\Gamma(1/\alpha_i)}\exp\left(-\frac{|e|^{\alpha_i}}{\beta_i^{\alpha_i}}\right);\ i=1,2,3. \tag{12}$$

where: $G_{\alpha_i,\beta_i}$: generalized Gaussian kernel; $\phi \in [0,1]$: mixture coefficient; $\Gamma$: Gamma function.

*Remark 2*: When $\kappa = 1$, the GMMEEF criterion is simplified into the GMMC criterion [15]; when $\kappa = 0$, the GMMEEF criterion is simplified into the GMEE criterion [9,13].

*Remark 3*: When $\alpha_1 = \alpha_2 = \alpha_3 = 2$, the GMMEEF criterion is simplified into the combination of the maximum mixture correntropy (MMC) criterion [14] and the MEE criterion [5].

*B. Derivation of GMMEEF-AUKF Algorithm*

*Time Update*

Consider a random variable **u** with a mean value $\hat{\mathbf{u}}$ and covariance matrix **P**. According to [25], **u** is calculated as follows:
$$\psi_{t-1|t-1}^a = \begin{cases} \hat{\mathbf{u}}_{t-1|t-1} & ; a=0 \\ \hat{\mathbf{u}}_{t-1|t-1} + \left(\sqrt{(n+\mu)\mathbf{P}_{t-1|t-1}}\right)_a & ; a=1,2,..,n \\ \hat{\mathbf{u}}_{t-1|t-1} - \left(\sqrt{(n+\mu)\mathbf{P}_{t-1|t-1}}\right)_{a-n} & ; a=a+1,..,2n \end{cases} \tag{13}$$

where $\left(\sqrt{(n+\mu)\mathbf{P}_{t-1|t-1}}\right)_a$ represents the *a* column vector of the matrix $\sqrt{(n+\mu)\mathbf{P}_{t-1|t-1}}$; typically obtained by performing a Cholesky decomposition on the original matrix; *α*: proportional correction factor *λ*: free coefficient; and $\mu = \alpha^2(n+\lambda) - n$.

Then, $\hat{\mathbf{u}}_{t|t-1}$ and $\mathbf{P}_{t|t-1}$ represents the prior state estimation and the prior state error covariance matrix are computed:
$$\hat{\mathbf{u}}_{t|t-1} = \sum_{a=0}^{2n} \rho_b^a \mathbf{f}\left(\psi_{t-1|t-1}^a\right) \tag{14}$$
$$\mathbf{P}_{t|t-1} = \sum_{a=0}^{2n} \rho_c^a \left(\mathbf{f}\left(\psi_{t-1|t-1}^a\right) - \hat{\mathbf{u}}_{t|t-1}\right)\left(\mathbf{f}\left(\psi_{t-1|t-1}^a\right) - \hat{\mathbf{u}}_{t|t-1}\right)^T + \mathbf{Q}_{t-1} \tag{15}$$

where: the variance weight $\rho_c^a$ and mean weight $\rho_b^a$ of Sigma points with $\rho_b^a = \rho_c^a = 1/(2n+2\mu)$ while $a \neq 0$; $\rho_b^0 = \mu/(n+\mu)$; and $\rho_c^0 = \mu/(n+\mu) + (1-\alpha^2+\beta)$.

*Measurement Update*

Continue to perform UT transformation on $\hat{\mathbf{u}}_{t|t-1}$ and $\mathbf{P}_{t|t-1}$ to obtain *2n+1* sigma point. Then, $\mathbf{P}_{uv,t}$ and $\hat{\mathbf{v}}_{t|t-1}$ the cross-covariance matrix and prior mean of measurement, respectively, are calculated by the following formulas:
$$\psi_{t|t-1}^a = \begin{cases} \hat{\mathbf{u}}_{t|t-1} & ; a=0 \\ \hat{\mathbf{u}}_{t|t-1} + \left(\sqrt{(n+\mu)\mathbf{P}_{t|t-1}}\right)_a & ; a=1,2,..,n \\ \hat{\mathbf{u}}_{t|t-1} - \left(\sqrt{(n+\mu)\mathbf{P}_{t|t-1}}\right)_{a-n} & ; a=n+1,..,2n \end{cases} \tag{16}$$

$$\hat{\mathbf{v}}_{t|t-1} = \sum_{a=0}^{2n} \rho_b^a \mathbf{g}_i(\psi_{t|t-1}^a) \tag{17}$$

$$\mathbf{P}_{uv,t|t-1} = \sum_{a=0}^{2n} \rho_c^a \left(\psi_{t|t-1}^a - \hat{\mathbf{u}}_{t|t-1}\right)\left(\mathbf{g}_i(\psi_{t|t-1}^a) - \hat{\mathbf{v}}_{t|t-1}\right)^T \tag{18}$$

$$\mathbf{P}_{vv,t|t-1} = \sum_{a=0}^{2n} \rho_c^a \left(\psi_{t|t-1}^a - \hat{\mathbf{u}}_{t|t-1}\right)\left(\mathbf{g}_i(\psi_{t|t-1}^a) - \hat{\mathbf{v}}_{t|t-1}\right)^T + \mathbf{R}_t \tag{19}$$

Similar to [8-15], to complete the measurement update process a dual noise model consisting of measurement variables and state variables is constructed. A measurement slope matrix can be defined as:

$$\mathbf{U}_t = \left(\mathbf{P}_{t|t-1}^{-1}\mathbf{P}_{uv,t}\right)^T \qquad (20)$$

Then, measurement (1) can be computed as follows:

$$\mathbf{v}_t = \hat{\mathbf{v}}_{t|t-1} + \mathbf{U}_t\left(\mathbf{u}_t - \hat{\mathbf{u}}_{t|t-1}\right) + \mathbf{r}_t \qquad (21)$$

Here, the prior state estimation error is described as:

$$\wp_t = \mathbf{u}_t - \hat{\mathbf{u}}_{t|t-1} \qquad (22)$$

Combining (21) and (22) yields:

$$\begin{bmatrix} \hat{\mathbf{u}}_{t|t-1} \\ \mathbf{v}_t - \hat{\mathbf{v}}_{t|t-1} + \mathbf{U}_t\hat{\mathbf{u}}_{t|t-1} \end{bmatrix} = \begin{bmatrix} \mathbf{I} \\ \mathbf{U}_t \end{bmatrix} \mathbf{u}_t + \boldsymbol{\chi}_t \qquad (23)$$

where $\mathbf{I}$: unity matrix; $\boldsymbol{\chi}_t = \begin{bmatrix} -\wp_t & \mathbf{r}_t \end{bmatrix}^T$ and

$$E\left[\boldsymbol{\chi}_t\boldsymbol{\chi}_t^T\right] = \begin{bmatrix} \mathbf{P}_{t|t-1} & 0 \\ 0 & \mathbf{R}_t \end{bmatrix} = \begin{bmatrix} \mathbf{B}_{P,t|t-1}\mathbf{B}_{P,t|t-1}^T & 0 \\ 0 & \mathbf{B}_{R,t}\mathbf{B}_{R,t}^T \end{bmatrix} = \mathbf{B}_t\mathbf{B}_t^T \qquad (24)$$

where $\mathbf{B}_{P,t|t-1} \in \mathbb{R}^{n\times n}$ and $\mathbf{B}_{R,t} \in \mathbb{R}^{m\times m}$: the Cholesky decomposition factors of $\mathbf{P}_{t/t-1}$ and $\mathbf{R}_t$, respectively. It should be noted that here the covariance matrix $E\left[\boldsymbol{\chi}_t\boldsymbol{\chi}_t^T\right]$ of the augmented error has been assumed to satisfy the positive definite condition [8-15]. Besides, some numerical stability enhancement solutions to satisfy the Cholesky decomposition condition have been introduced in [26-28].

Continue multiplying both sides of (23) by $\mathbf{B}_t^{-1}$, obtaining:

$$\mathbf{L}_t = \mathbf{D}_t\mathbf{u}_t + \mathbf{e}_t \qquad (25)$$

$$\mathbf{L}_t = \mathbf{B}_t^{-1}\begin{bmatrix} \hat{\mathbf{u}}_{t|t-1} \\ \mathbf{v}_t - \hat{\mathbf{v}}_{t|t-1} + \tilde{\mathbf{U}}_t\hat{\mathbf{u}}_{t|t-1} \end{bmatrix} = \begin{bmatrix} l_{1,t}, l_{2,t}, ..., l_{n+m,t} \end{bmatrix}^T \qquad (26)$$

$$\mathbf{D}_t = \mathbf{B}_t^{-1}\begin{bmatrix} \mathbf{I} \\ \tilde{\mathbf{U}}_t \end{bmatrix} = \begin{bmatrix} \mathbf{d}_{1,t}^T, \mathbf{d}_{2,t}^T, ..., \mathbf{d}_{n+m,t}^T \end{bmatrix}^T \in \mathbb{R}^{(n+m)\times n} \qquad (27)$$

$$\mathbf{e}_t = \mathbf{B}_t^{-1}\begin{bmatrix} -\wp_t \\ \mathbf{r}_t \end{bmatrix} = \begin{bmatrix} e_{1,t}, ..., e_{n+m,t} \end{bmatrix}^T \in \mathbb{R}^{(n+m)\times 1} \qquad (28)$$

Note that $E\left[\mathbf{e}_t\mathbf{e}_t^T\right] = \mathbf{I}_N$ and the residual error $\mathbf{e}_t$ is white. According to the proposed GMMEEF criterion, the cost function of GMMEEF-UKF can be defined by:

$$J_{GMMEEF}(u_t) = \kappa\sum_{i=1}^{N}\left[\phi G_{\alpha_1,\beta_1}(e_i) + (1-\phi)G_{\alpha_2,\beta_2}(e_i)\right] + (1-\kappa)\sum_{i=1}^{N}\sum_{j=1}^{N} G_{\alpha_3,\beta_3}(e_j - e_i) \qquad (29)$$

where: $e_{i,t} = l_{i,t} - \mathbf{d}_{i,t}\mathbf{u}_t$; $l_{i,t}$ represent the $i^{th}$ element of $\mathbf{e}_t$ and $\mathbf{L}_t$ respectively; $\mathbf{d}_{i,t}$ represent the $i^{th}$ row of $\mathbf{D}_t$. The optimal estimate can be obtained by calculating $\hat{u}_t = \arg\max[J(u_t)]$. Taking the derivative of $J(u_t)$ with respect to $u_t$ be zero:

$$\frac{\partial J_{GMMEEF}}{\partial u_t} = \kappa\sum_{i=1}^{N}\left(\phi_j\frac{\alpha_j}{\beta_j^{\alpha_j}}G_{\alpha_j,\beta_j}(e_{t,j})|e_{t,j}|^{\alpha_j-2}\right)\mathbf{d}_{t,j}^T e_{t,j}$$
$$+ (1-\kappa)\sum_{i=1}^{N}\sum_{j=1}^{N}G_{\alpha_3,\beta_3}(e_{t,j} - e_{t,i})|e_{t,j} - e_{t,i}|^{\alpha_j-2}\mathbf{d}_{t,j}^T e_{t,j}$$
$$= 0 \qquad (30)$$

Simplifying Eq. (30) as follows:

$$\frac{\partial J_{GMMEEF}}{\partial u_t} = \kappa\mathbf{D}_t^T\boldsymbol{\Lambda}_t\mathbf{e}_t + (1-\kappa)\mathbf{D}_t^T(\boldsymbol{\Phi}_t - \boldsymbol{\Xi}_t)\mathbf{e}_t = 0 \qquad (31)$$

$$\boldsymbol{\Lambda}_t = diag\begin{bmatrix} \sum_{j=1}^{2}\left(\phi_j\frac{\alpha_j}{\beta_j^{\alpha_j}}G_{\alpha_j,\beta_j}(e_{t,1})|e_{t,1}|^{\alpha_j-2}\right), ... \\ ..., \sum_{j=1}^{2}\left(\phi_j\frac{\alpha_j}{\beta_j^{\alpha_j}}G_{\alpha_j,\beta_j}(e_{t,N})|e_{t,N}|^{\alpha_j-2}\right) \end{bmatrix} \qquad (32)$$

$$\boldsymbol{\Phi}_t = diag\begin{bmatrix} \sum_{j=1}^{N}G_{\alpha_3,\beta_3}(e_{t,1} - e_{t,j})|e_{t,1} - e_{t,j}|^{\alpha_j-2}, ... \\ ..., \sum_{j=1}^{N}G_{\alpha_3,\beta_3}(e_{t,N} - e_{t,j})|e_{t,N} - e_{t,j}|^{\alpha_j-2} \end{bmatrix} \qquad (33)$$

$$[\boldsymbol{\Xi}_t]_{ij} = G_{\alpha_3,\beta_3}(e_{t,j} - e_{t,i})|e_{t,j} - e_{t,i}|^{\alpha_j-2} \qquad (34)$$

with $[\boldsymbol{\Xi}_t]_{ij}$ present the element in row $j$ and column $k$ of $\boldsymbol{\Xi}_t$. Similar to the derivation in [8-15], the results as follows:

$$\mathbf{u}_t = (\mathbf{D}_t^T\boldsymbol{\Omega}_t\mathbf{D}_t)^{-1}\mathbf{D}_t^T\boldsymbol{\Omega}_t\mathbf{L}_t \qquad (35)$$

With $\boldsymbol{\Omega}_t = \kappa\boldsymbol{\Lambda}_t + (1-\kappa)(\boldsymbol{\Phi}_t - \boldsymbol{\Xi}_t)$ and the matrix $\boldsymbol{\Omega}_t$ can also be described as follows:

$$\boldsymbol{\Omega}_t = \begin{bmatrix} \boldsymbol{\Omega}_{uu,t} & \boldsymbol{\Omega}_{vu,t} \\ \boldsymbol{\Omega}_{uv,t} & \boldsymbol{\Omega}_{vv,t} \end{bmatrix} \qquad (36)$$

where: $\boldsymbol{\Omega}_{uu,t} \in \mathbb{R}^{n\times n}; \boldsymbol{\Omega}_{uv,t} \in \mathbb{R}^{m\times n}; \boldsymbol{\Omega}_{uv,t} \in \mathbb{R}^{n\times m}; \boldsymbol{\Omega}_{vv,t} \in \mathbb{R}^{m\times m}$

Observing Eq.(35), it can be seen that $\mathbf{u}_t$ as $\mathbf{u}_t = h(\mathbf{u}_t)$, which can be obtained when using fixed point iterative. According to (25,36) and the matrix inverse lemma, the Eq. (35) can be rewritten as:

$$\hat{\mathbf{u}}_{t|t} = \hat{\mathbf{u}}_{t|t-1} + \bar{\mathbf{K}}_t(\mathbf{v}_t - \hat{\mathbf{v}}_{t|t-1}) \qquad (37)$$

where the gain matrix $\bar{\mathbf{K}}_t$ as follows:

$$\bar{\mathbf{K}}_t = (\bar{\mathbf{P}}_{uu,t} + \mathbf{U}_t^T\bar{\mathbf{P}}_{uv,t} + \bar{\mathbf{P}}_{vu,t}\mathbf{U}_t + \mathbf{U}_t^T\bar{\mathbf{R}}_{vv,t}\mathbf{U}_t)^{-1}(\bar{\mathbf{P}}_{vu,t} + \mathbf{U}_t^T\bar{\mathbf{R}}_{vv,t}) \qquad (38)$$

$$\begin{cases} \bar{\mathbf{P}}_{uu,t} = (\mathbf{B}_{P,t|t-1}^{-1})^T\boldsymbol{\Omega}_{uu,t}\mathbf{B}_{P,t|t-1}^{-1} \\ \bar{\mathbf{P}}_{uv,t} = (\mathbf{B}_{R,t}^{-1})^T\boldsymbol{\Omega}_{uv,t}\mathbf{B}_{P,t|t-1}^{-1} \\ \bar{\mathbf{P}}_{vu,t} = (\mathbf{B}_{P,t|t-1}^{-1})^T\boldsymbol{\Omega}_{vu,t}\mathbf{B}_{R,t}^{-1} \\ \bar{\mathbf{R}}_{vv,t} = (\mathbf{B}_{R,t}^{-1})^T\boldsymbol{\Omega}_{vv,t}\mathbf{B}_{R,t}^{-1} \end{cases} \qquad (39)$$

The state covariance matrix $\mathbf{P}_{t|t} = E\left[(\mathbf{u}_t - \hat{\mathbf{u}}_{t|t})(\mathbf{u}_t - \hat{\mathbf{u}}_{t|t})^T\right]$ is computed by:

$$\mathbf{P}_{t|t} = (\mathbf{I}_n - \bar{\mathbf{K}}_t\mathbf{U}_t)\mathbf{P}_{t|t-1}(\mathbf{I}_n - \bar{\mathbf{K}}_t\mathbf{U}_t)^T + \bar{\mathbf{K}}_t\mathbf{R}_t\bar{\mathbf{K}}_t^T \qquad (40)$$

To satisfy the condition for performing Cholesky decomposition when computing the square root of $\mathbf{P}_{t-1|t-1}$ in Eq.(13), the $QR$ decomposition method to compute the square root of the matrix is applied [6,28]. Eq.(40) is rewritten as follows:

$$\mathbf{P}_{t|t} = \mathbf{S}_{t|t}\mathbf{S}_{t|t}^T \qquad (41)$$

$$\mathbf{S}_{t|t} = \left[\left(\mathbf{I}_{n\times n} - \bar{\mathbf{K}}_{t|t}\mathbf{U}_{t|t}\right)\mathbf{B}_{P,t|t} \quad \bar{\mathbf{K}}_{t|t}\mathbf{B}_{R,t|t}\right] \qquad (42)$$

Performing the $QR$ decomposition on $\mathbf{S}_{t|t}^T$, obtained:

$$\mathbf{S}_{t|t}^T = \mathbf{T}_{t|t}\begin{bmatrix} \mathbf{A}_{t|t} \\ \mathbf{0}_{m\times n} \end{bmatrix} \qquad (43)$$

where: $\mathbf{T}_{t|t} \in \mathbb{R}^{(n+m)\times(n+m)}$: orthogonal matrix; $\mathbf{A}_{t|t} \in \mathbb{R}^{n\times n}$: upper triangular matrix. Combining Eq.(41) and Eq.(43), received:

$$\mathbf{P}_{t|t} = \mathbf{A}_{t|t}^T \mathbf{A}_{t|t} \quad (44)$$

Note that the square matrix $\mathbf{A}_{t|t}^T$ is the square root matrix of $\mathbf{P}_{t|t}$. Since the *QR* decomposition has no restrictions on the original matrix, Eq.(43) is numerically stable.

It can be seen that the covariance matrices $\mathbf{Q}_t$ and $\mathbf{R}_t$ in Kalman filters play a particularly important role, greatly impacting the estimated accuracy. However, under the dynamic changes of the system, these matrices are deeply affected. Therefore, at each time *t*, these two matrices need to be dynamically estimated. According to [21,23,24], $\mathbf{Q}_t$ and $\mathbf{R}_t$ can be estimated as follows:

$$\hat{\mathbf{Q}}_t = (1-\theta_t)\hat{\mathbf{Q}}_{t-1} + \theta_t [\mathbf{K}_t \tilde{\mathbf{v}}_t \tilde{\mathbf{v}}_t^T \mathbf{K}_t^T + \mathbf{P}_{t|t} - \ldots \\ - \sum_{a=0}^{2n} \rho_c^a \left( \mathbf{f}(\psi_{t-1|t-1}^a) - \hat{\mathbf{u}}_{t|t-1} \right)\left( \mathbf{f}(\psi_{t-1|t-1}^a) - \hat{\mathbf{u}}_{t|t-1} \right)^T ] \quad (45)$$

$$\hat{\mathbf{R}}_{t+1} = (1-\theta_t)\hat{\mathbf{R}}_t + \theta_t [\tilde{\mathbf{v}}_t \tilde{\mathbf{v}}_t^T + \ldots \\ + \sum_{a=0}^{2n} \rho_c^a \left( \psi_{t|t-1}^a - \hat{\mathbf{u}}_{t|t-1} \right)\left( \mathbf{g}_t(\psi_{t|t-1}^a) - \hat{\mathbf{v}}_{t|t-1} \right)^T ] \quad (46)$$

where $\theta_t = (1-s)/(1-s^{t+1})$; *s*: the forgetting factor; $\tilde{\mathbf{v}}_t = \mathbf{v}_t - \tilde{\mathbf{v}}_{t|t-1}$: innovation vector.

Combining Eq.(15) and Eq.(19), Eq.(45,46) can be rewritten as follows:

$$\hat{\mathbf{Q}}_t = (1-\theta_t)\hat{\mathbf{Q}}_{t-1} + \theta_t \left[ \mathbf{K}_t \tilde{\mathbf{v}}_t \tilde{\mathbf{v}}_t^T \mathbf{K}_t^T + \mathbf{P}_{t|t} - \mathbf{P}_{t|t-1} + \hat{\mathbf{Q}}_{t-1} \right] \quad (47)$$

$$\hat{\mathbf{R}}_{t+1} = (1-\theta_t)\hat{\mathbf{R}}_t + \theta_t \left[ \tilde{\mathbf{v}}_t \tilde{\mathbf{v}}_t^T - \mathbf{P}_{vv,t|t-1} + \hat{\mathbf{R}}_t \right] \quad (48)$$

---

**Algorithm 1** Pseudocode of the GMMEEF-AUKF

*Step 1:* Set initial $\hat{\mathbf{u}}_{0|0}$; $\mathbf{P}_{0|0}$; $\hat{\mathbf{Q}}_0$; $\hat{\mathbf{R}}_1$; $\alpha$; $\beta$; $\alpha_1$; $\alpha_2$; $\alpha_3$; $\beta_1$; $\beta_2$; $\beta_3$; $\theta$; $\kappa$; $\phi$; $\delta$ (threshold)

*For:* $t = 1, 2, 3, \ldots$

*Step 2:* Calculating $\hat{\mathbf{u}}_{t|t-1}$; $\mathbf{P}_{t|t-1}$; $\hat{\mathbf{v}}_{t|t-1}$; $\mathbf{P}_{uv,t|t-1}$ through Eq.(14), (15), (17), (18)
Calculating $\mathbf{U}_t$ through Eq. (20)

*Step 3:* Calculating $\mathbf{B}_{P,t|t-1}$; $\mathbf{B}_{R,t}$ through Eq.(24)
Calculating $\mathbf{L}_t$; $\mathbf{D}_t$ through Eq.(26), (27)

*Step 4:* Set $k=1$; $\hat{\mathbf{u}}_{t|t}^0 = \hat{\mathbf{u}}_{t|t-1}$
Calculating $\mathbf{e}_t^k = \mathbf{L}_t - \mathbf{D}_t \hat{\mathbf{u}}_{t|t}^{k-1}$
Calculating $\overline{\mathbf{K}}_t$ through Eq.(38)
Calculating $\hat{\mathbf{u}}_{t|t}^k = \hat{\mathbf{u}}_{t|t-1} + \overline{\mathbf{K}}_t (\mathbf{v}_t - \hat{\mathbf{v}}_{t|t-1})$
If $\left\| \hat{\mathbf{u}}_{t|t}^k - \hat{\mathbf{u}}_{t|t}^{k-1} \right\| / \left\| \hat{\mathbf{u}}_{t|t}^{k-1} \right\| \le \delta$ hold, set $\hat{\mathbf{u}}_t = \hat{\mathbf{u}}_{t|t}^k$ and go to *step 5*.
Otherwise, set $k = k+1$, and go back calculating $\mathbf{e}_t^k$

*Step 5:* Update $\mathbf{P}_{t|t}$, $\hat{\mathbf{Q}}_t$, $\hat{\mathbf{R}}_{t+1}$ through Eq.(44),(47),(48)
Revise $\hat{\mathbf{Q}}_t$, $\hat{\mathbf{R}}_{t+1}$ through Eq.(49),(50)

*End for*

---

Since $\mathbf{q}_t$; $\mathbf{r}_t$ are uncorrelated noises, so consider $\mathbf{Q}_t$ and $\mathbf{R}_t$ as diagonal matrices. To satisfy the conditions of Cholesky's decomposition, it can be described as follows:

$$\hat{\mathbf{Q}}_t = \sqrt{diag(\hat{\mathbf{Q}}_t \hat{\mathbf{Q}}_t^T)} \quad (49)$$

$$\hat{\mathbf{R}}_{t+1} = \sqrt{diag(\hat{\mathbf{R}}_{t+1} \hat{\mathbf{R}}_{t+1}^T)} \quad (50)$$

The estimates of $\mathbf{Q}_t$ and $\mathbf{R}_{t+1}$ are positive definite. In summary of the above analysis, the pseudocode of the GMMEF-AUKF is shown in algorithm 1.

## IV. ISGA-GMMEEF-AUKF ALGORITHM

It can be observed that too many coefficients must to choose the value in GMMEEF-AUKF: the shape coefficients of the kernel ($\alpha_1, \alpha_2, \alpha_3, \beta_1, \beta_2, \beta_3$) in the GMMEEF, the scale coefficients that impact the selection of sigma points ($\alpha, \beta$) in the UT, and the update coefficient ($\theta$) of the noise covariance matrices. These coefficients affect the quality of the UKF algorithm using the correntropy and entropy criterion [9,13,15-17]. To solve this problem, meta-heuristic algorithms are often utilized to automatically calculate and select the best values of these coefficients that fit with the PS [17-19]. In this paper, the ISGA is designed. The flow chart of the ISGA-GMMEEF-AUKF algorithm is given in Figure 1.

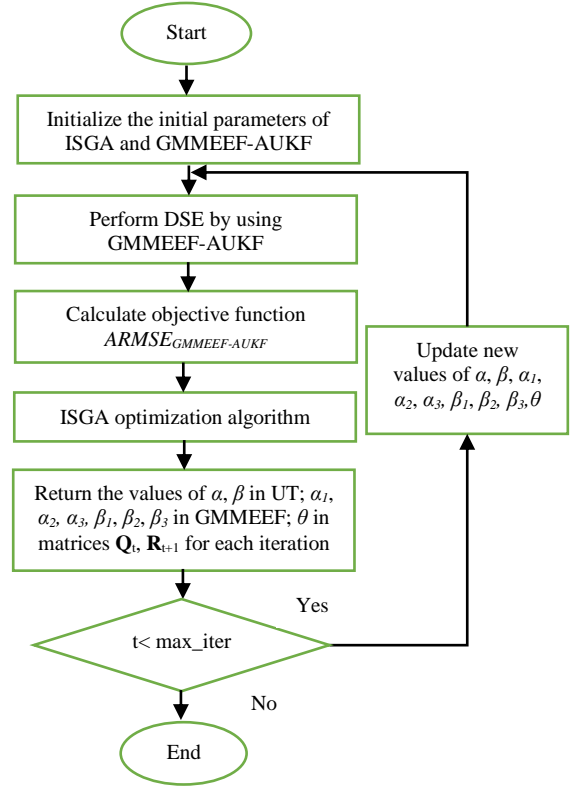

Fig. 1. Flowchart of the ISGA-GMMEEF-AUKF algorithm

Snow Geese is famous for harsh long migration. To complete that migration, Snow Geese has optimized energy through the layout of the flying squad (herringbone shape or straight line shape) and flight height. Based on this inspiration, an optimal algorithm was introduced in 2024 [29]. Snow Geese algorithm (SGA) optimized based on population, each individual (agent) will represent an optimal solution $TF(X_i)$ with the position $X_i$. Similar to other meta-heuristic algorithms, SGA's optimal process is conducted in two phases: exploration and exploitation.

These two phases are converted through parameter $\omega$ ($M$: Maximum iterations).

$$\omega = \frac{2\pi t}{M} \quad (51)$$

- In the exploration phase (flying in the form of herringbone): in the flying squad, each Snow Geese layer will have a different way of updating the position. Specifically:

+ For the first one-fifth of individuals, each agent has updated the position according to Eq.(52):

$$X_i^{t+1} = X_i^t + b(X_b^t - X_i^t) + V_i^{t+1} \quad (52)$$

+ For the agent in the midsection, each agent has updated the position according to Eq.(53):

$$X_i^{t+1} = X_i^t + b(X_b^t - X_i^t) - d(X_c^t - X_i^t) + V_i^{t+1} \quad (53)$$

+ For the remaining agent, each agent has updated the position according to Eq.(54):

$$X_i^{t+1} = X_i^t + b(X_b^t - X_i^t) + d(X_c^t - X_i^t) - \eta(X_n^t + X_i^t) + V_i^{t+1} \quad (54)$$

$$\begin{cases} X_c^t = \dfrac{\sum_{i=1}^n X_i^t f(X_i^t)}{n \sum_{i=1}^n f(X_i^t)} \\ V_i^{t+1} = \dfrac{4t}{Me^{\frac{4t}{M}}} V_i^t + X_b^t - X_i^t - \dfrac{1,29.(V_i^t)^2.10^{-2}.\sin(\omega)}{2} \end{cases} \quad (55)$$

with $b = 4rand() - 2$ ; $d = 3rand() - 1.5$ ; $\eta = 2rand() - 1$ and $X_b^t$: The best position; $X_i^t$: The position of an individual Snow Geese; $V_i^{t+1}$: The next generation velocity; $V_i^t$: The current velocity; $n$: Population size.

- In the exploitation phase (flying in the form of a straight line): each agent is updated the position according to Eq.(56):

$$X_i^{t+1} = \begin{cases} X_i^t + (X_i^t - X_b^t)r & ; r > 0.5 \\ X_i^t + (X_i^t - X_b^t)r \oplus Brownian(d) & ; r \leq 0.5 \end{cases} \quad (56)$$

where: $r$: random number; $Brownian(d)$: Brownian motion

In addition, the BAT algorithm was introduced in 2010 [30]. The highlight of BAT is its echolocation capability. The mathematical description of this process is as follows:

$$f_i = f_{min} + (f_{max} - f_{min}).\xi \quad (57)$$

$$v_i^{t+1} = v_i^t + (x_i^t - x_*)f_i \quad (58)$$

$$x_i^{t+1} = x_i^t + v_i^t \quad (59)$$

where: $x_i$: the position of bats; $x_*$: the best position of bats; $v_i$: the velocity of bats; $f_i$; $f_{min}$; $f_{max}$: current frequency, maximum frequency, and minimum frequency of waves, respectively; $\xi$: a random number in [0,1].

To increase the optimal searchability, the exploitation phase in BAT is utilized to replace the exploitation phase in SGA. From the above analysis, to enhance the GMMEEF-AUKF performance, ISGA is utilized to find the optimal value of coefficients. First, GMMEEF-AUKF performs state estimation and gives an expression for evaluating the average root mean square error (*ARMSE*) of the estimation. Next, ISGA considers *ARMSE* as an objective function (OF) and finds the optimal value of the above coefficients so that *ARMSE* reaches the minimum value. Here, it should be noted that $TF(\mathbf{X}_i) = OF_{ARMSE}$ and $\mathbf{X}_i = [\alpha, \beta, \alpha_1, \alpha_2, \alpha_3, \beta_1, \beta_2, \beta_3, \theta]$. The pseudocode of ISGA-GMMEEF-AUKF is given in algorithm 2.

---

**Algorithm 2** Pseudocode of the ISGA-GMMEEF-AUKF

**Step 1:** Set initial $\hat{\mathbf{u}}_{0|0}$ ; $\mathbf{P}_{0|0}$ ; $\hat{\mathbf{Q}}_0$ ; $\hat{\mathbf{R}}_1$ ; $\kappa$; $\phi$; $\delta$, $n$, $M$; $f_{min}$; $f_{max}$
**For:** $t = 1, 2, 3, \ldots$
**Step 2:** Calculating the factors the same as **Algorithm 1**
**Step 3:** Calculating $OF_{ARMSE}(GMMEEF - AUKF)$
  Calculate the fitness value $TF(\mathbf{X}_i)$ of the individuals
  $(TF(\mathbf{X}_i) = OF_{ARMSE})$
  Record the best value $TF(\mathbf{X}_b)$ and the best position $\mathbf{X}_b$
**Step 4:** While $k < M$ do Computing $\omega$ by Eq. (51)
    If $\omega < \pi$ then (exploration phase)
   Update position according to Eq. (52,53,54)
    Else (exploitation phase)
   Update position according to Eq. (57,58,59)
    End if
   Calculation of next-generation fit values
    For $i=1$; $i \leq n$ do
    If $TF(\mathbf{X}_b^t) > TF(\mathbf{X}_b^{t+1})$ then
    Update the best position $\mathbf{X}_b^{t+1} = \mathbf{X}_b^{t+1}$
     Else Update the best position $\mathbf{X}_b^{t+1} = \mathbf{X}_b^t$
    End if
   End for
   $k = k + 1$
  End while
**Step 5:** Updating $\alpha, \beta, \alpha_1, \alpha_2, \alpha_3, \beta_1, \beta_2, \beta_3, \theta$
  and go back to **step 2**
**End For**

---

## V. PROOF OF CONVERGENCE

In this section, the convergent proof of the estimated value obtained through the fixed-point iteration method based on the references [8,31,32] is provided. Similar to [8], suppose the kernel size satisfies the condition: $\alpha_1 = \alpha_2 = c\alpha_3$ (where c, $\alpha_1$, $\alpha_2$, $\alpha_3$: nonnegative constant). First, $\mathbf{u}_t$ in Eq. (35) can be rewritten:

$$\mathbf{u}_t = \mathbf{W}_{MM}^{-1} \mathbf{W}_{DM} \quad (60)$$

$$\mathbf{W}_{MM} = \kappa \sum_{j=1}^{N} \left[ \phi G_{\alpha_1, \beta_1}(e_{j,t}) + (1-\phi) G_{\alpha_2, \beta_2}(e_{j,t}) \right] \mathbf{d}_{j,t} \mathbf{d}_{j,t}^T + (1-\kappa) \sum_{i=1}^{N} \sum_{j=1}^{N} G_{\alpha_3, \beta_3}(e_{j,t} - e_{i,t}) \mathbf{\Pi}_1 \quad (61)$$

$$\mathbf{W}_{DM} = \kappa \sum_{j=1}^{N} \left[ \phi G_{\alpha_1, \beta_1}(e_{j,t}) + (1-\phi) G_{\alpha_2, \beta_2}(e_{j,t}) \right] \mathbf{d}_{j,t} l_{j,t} + (1-\kappa) \sum_{i=1}^{N} \sum_{j=1}^{N} G_{\alpha_3, \beta_3}(e_{j,t} - e_{i,t}) \mathbf{\Pi}_2 \quad (62)$$

$$\mathbf{\Pi}_1 = \left[ \mathbf{d}_{i,t} - \mathbf{d}_{j,t} \right] \left[ \mathbf{d}_{i,t} - \mathbf{d}_{j,t} \right]^T \quad (63)$$

$$\mathbf{\Pi}_2 = \left[ l_{i,t} - l_{j,t} \right] \left[ \mathbf{d}_{i,t} - \mathbf{d}_{j,t} \right]^T \quad (64)$$

Based on [29,30], it is obtained:

$$\left\| \mathbf{W}_{\mathbf{MM}}^{-1} \mathbf{W}_{\mathbf{DM}} \right\|_1 \leq h(\mathbf{u}_t) = \frac{\sqrt{n}\left(\kappa \mathbf{\Pi}_3 + (1-\kappa)\mathbf{\Pi}_4\right)}{W_{\min}\left(\kappa \mathbf{\Pi}_5 + (1-\kappa)\mathbf{\Pi}_6\right)} \quad (65)$$

$$\mathbf{\Pi}_3 = \sum_{j=1}^{N} \left\| \mathbf{d}_{j,t} \right\|_1 \left| l_{j,t} \right| \quad (66)$$

$$\mathbf{\Pi}_4 = \sum_{i=1}^{N}\sum_{j=1}^{N} \left| l_{i,t} - l_{j,t} \right| \left\| \mathbf{d}_{i,t} - \mathbf{d}_{j,t} \right\|_1 \quad (67)$$

$$\mathbf{\Pi}_5 = \sum_{j=1}^{N} \left[ \begin{array}{l} \phi G_{\alpha_1,\beta_1}\left(s \left\| \mathbf{d}_{j,t} \right\|_1 + \left| l_{j,t} \right|\right) \mathbf{d}_{j,t} \mathbf{d}_{j,t}^T + \\ +(1-\phi) G_{\alpha_2,\beta_2}\left(s \left\| \mathbf{d}_{j,t} \right\|_1 + \left| l_{j,t} \right|\right) \mathbf{d}_{j,t} \mathbf{d}_{j,t}^T \end{array} \right] \quad (68)$$

$$\mathbf{\Pi}_6 = \sum_{i=1}^{N}\sum_{j=1}^{N} G_{\alpha_3,\beta_3}\left(s \left\| \mathbf{d}_{i,t} - \mathbf{d}_{j,t} \right\|_1 + \left| l_{i,t} - l_{j,t} \right|\right) \mathbf{\Pi}_1 \quad (69)$$

where: $W_{\min}$: the minimum eigenvalue of $\mathbf{W}_{\mathbf{MM}}$; $s = y_1(\alpha_3)$.

Additionally, the Jacobian matrix of $h(\mathbf{u}_t)$ about $\mathbf{u}_t$ is calculated by:

$$\nabla_{\hat{\mathbf{u}}_t} h(\mathbf{u}_t) = \frac{\partial}{\partial \mathbf{u}_t} \mathbf{W}_{\mathbf{MM}}^{-1} \mathbf{W}_{\mathbf{DM}}$$

$$= -\mathbf{W}_{\mathbf{MM}}^{-1} \left[ \begin{array}{l} \frac{\kappa}{c^2\alpha_3^2} \sum_{j=1}^{N} e_{j,t} \mathbf{d}_{j,t}^s \left( \begin{array}{l} \phi G_{\alpha_1,\beta_1}(e_{j,t}) \\ +(1-\phi) G_{\alpha_2,\beta_2}(e_{j,t}) \end{array} \right) \times \mathbf{d}_{j,t} \mathbf{d}_{j,t}^T \\ + \frac{1-\kappa}{\alpha_3^2} \sum_{i=1}^{N}\sum_{j=1}^{N} \Psi_1 \mathbf{\Pi}_1 \end{array} \right] h(\mathbf{u}_t)$$

$$+ \mathbf{W}_{\mathbf{DM}}^{-1} \left[ \begin{array}{l} \frac{\kappa}{c^2\alpha_3^2} \sum_{j=1}^{N} e_{j,t} \mathbf{d}_{j,t}^s \left( \begin{array}{l} \phi G_{\alpha_1,\beta_1}(e_{j,t}) \\ +(1-\phi) G_{\alpha_2,\beta_2}(e_{j,t}) \end{array} \right) \times \mathbf{d}_{j,t} l_{j,t} \\ + \frac{1-\kappa}{\alpha_3^2} \sum_{i=1}^{N}\sum_{j=1}^{N} \Psi_1 \mathbf{\Pi}_2 \end{array} \right] \quad (70)$$

where: $\Psi_1 = \left[ e_{j,t} - e_{i,t} \right] \left[ \mathbf{d}_{j,t}^s - \mathbf{d}_{i,t}^s \right] G_{\alpha_3,\beta_3}(e_{j,t} - e_{i,t})$

According to the proof in [31,32], it is obtained:

$$\left\| \nabla_{\hat{\mathbf{u}}_t} h(\mathbf{u}_t) \right\|_1 \leq y_2(\alpha_3) = \frac{\sqrt{n}\left(\kappa \mathbf{\Pi}_7 + (1-\kappa)\mathbf{\Pi}_8\right)}{W_{\min}\left(c^2\alpha_3^2 \kappa \mathbf{\Pi}_5 + 2\alpha_3^2(1-\kappa)\mathbf{\Pi}_6\right)} \quad (71)$$

$$\mathbf{\Pi}_7 = \sum_{j=1}^{L} \left[ \left(s\left\|\mathbf{d}_{j,t}\right\|_1 + \left|l_{j,t}\right|\right) \left\|\mathbf{d}_{j,t}\right\|_1 \times \left(s\left\|\mathbf{d}_{j,t}\mathbf{d}_{j,t}^T\right\|_1 + \left\|l_{j,t}\mathbf{d}_{j,t}\right\|_1\right) \right] \quad (72)$$

$$\mathbf{\Pi}_8 = \sum_{i=1}^{N}\sum_{j=1}^{N} \left[ \begin{array}{l} \left(s\left\|\mathbf{d}_{i,t} - \mathbf{d}_{j,t}\right\|_1 + \left|l_{i,t} - l_{j,t}\right|\right)\left\|\mathbf{d}_{i,t} - \mathbf{d}_{j,t}\right\|_1 \times \\ \left(s\left\|(\mathbf{d}_{i,t} - \mathbf{d}_{j,t})(\mathbf{d}_{i,t} - \mathbf{d}_{j,t})^T\right\|_1 + \left|l_{i,t} - l_{j,t}\right|\left\|\mathbf{d}_{i,t} - \mathbf{d}_{j,t}\right\|_1\right) \end{array} \right] \quad (73)$$

Base on Theorem 1 in [8], If the kernel size value satisfies $\alpha_3 \geq \max\{\alpha^{\dagger 1}, \alpha^{\dagger 3}\}$ and

$$s > \frac{\sqrt{n}\left(\kappa \mathbf{\Pi}_3 + (1-\kappa)\mathbf{\Pi}_4\right)}{W_{\min}\left[\kappa \sum_{j=1}^{N} \mathbf{d}_{j,t}\mathbf{d}_{j,t}^T + (1-\kappa)\sum_{i=1}^{N}\sum_{j=1}^{N}\mathbf{\Pi}_1\right]} \quad (74)$$

where: $\alpha^{\dagger 1}, \alpha^{\dagger 3}$: the solutions of $y_1(\alpha_3)$ and $y_2(\alpha_3)$, respectively, received as follows (where: $\hbar = y_2(\alpha_3)$).

$$\begin{cases} \left\| h(\mathbf{u}_t) \right\|_1 \leq s \\ \left\| \nabla_{\hat{\mathbf{u}}_t} h(\mathbf{u}_t) \right\|_1 = \left\| \frac{\partial h(\mathbf{u}_t)}{\partial \mathbf{u}_t} \right\|_1 \leq \hbar \leq 1 \end{cases} \quad (75)$$

If $\alpha_3 \geq \max\{\alpha^{\dagger 1}, \alpha^{\dagger 3}\}$ and $\|\mathbf{u}_t\|_1 \leq s$, $\mathbf{u}_t$ is guaranteed to converges to a unique fixed-point in the range $\mathbf{u}_t \in \{\|\mathbf{u}_t\|_1 \leq s\}$.

Thus, the proposed algorithm obtains an estimated value that ensures convergence.

## VI. EXPERIMENTAL RESULTS

The PS parameters are shown in Table I ($k_i$: voltage amplitude; $k_v$: power injection; $k_p$: power flow, n: number of states, m: number of measurements). The initial conditions such as independent Monte Carlo experiments $D = 200$; total sample time T=60; scaling factor κ=0.5; mixture coefficients $\phi$=0.5; δ=10$^{-6}$; $\mathbf{P}_{0|0} = 10^{-2}$; $\mathbf{R}_0 = 10^{-2}\mathbf{I}_m$: measurement covariance matrices; $\mathbf{Q}_0 = 10^{-5}\mathbf{I}_n$: process covariance matrices. The data and the initial state $\mathbf{u}_0$ are used the same as the website [33] to test the system and $E\left[\hat{\mathbf{u}}_{0|0}\right] = \mathbf{u}_0$; $E\left[(\mathbf{u}_0 - \hat{\mathbf{u}}_{0|0})(\mathbf{u}_0 - \hat{\mathbf{u}}_{0|0})^T\right] = \mathbf{P}_{0|0}$. The value of coefficients in GMMEEF-AUKF is set as follows: $\theta = 0.5$; $\alpha = 10^{-2}$, $\beta = 1$, $\alpha_1 = 2.1$, $\alpha_2 = 2.1$, $\alpha_3 = 2.9$, $\beta_1 = 6.3$, $\beta_2 = 6.3$, $\beta_3 = 3.2$. It should be noted that the values of all the coefficients are selected based on references [13,16,18] and through testing.

The simulation program is run on a Core™ i7-5600U-CPU 2.60GHz computer. In this paper, the AUKF over ISGA (ISGA-AUKF) algorithm has been built to compare performance. ISGA-GMMEEF-AUKF and GMMEEF-AUKF are compared with UKF, AUKF, ISGA-AUKF, MCC-UKF [4,6], MEE-UKF [4,5], and MEEF-UKF [4] to increase convincingness. These algorithms estimate the voltage amplitude and phase on Bus-5 of IEEE systems.

TABLE I
PARAMETERS OF THE PS MODEL

| Parameters | 57-bus | 30-bus | 14-bus |
|---|---|---|---|
| n | 113 | 59 | 27 |
| m | 331 | 172 | 96 |
| $k_i$ | 80 | 41 | 20 |
| $k_v$ | 57 | 30 | 14 |
| $k_p$ | 57 | 30 | 14 |

This paper evaluates the amplitude and phase errors of the voltage using the *ARMSE* criterion:

$$ARMSE(\mathbf{V}_t) = \sqrt{\frac{1}{AD}\sum_{j=1}^{D}\left\|\hat{\mathbf{V}}_t^j - \mathbf{V}_t^j\right\|_2^2} \Big/ N \quad (76)$$

$$ARMSE(\mathbf{\varphi}_t) = \sqrt{\frac{1}{AD}\sum_{j=1}^{D}\left\|\hat{\mathbf{\varphi}}_t^j - \mathbf{\varphi}_t^j\right\|_2^2} \Big/ N \quad (77)$$

where: *N*: number of samples, *A*: number of buses, *D*: number of independent Monte-Carlo experiments; $\mathbf{V}_t^j$, $\hat{\mathbf{V}}_t^j$: actual value and the estimated value of voltage amplitude, $\mathbf{\varphi}_t^j$, $\hat{\mathbf{\varphi}}_t^j$: actual value and the estimated value of voltage phase respectively at the *j*$^{th}$ Monte-Carlo experiments.

## A. Performance Evaluation of The ISGA

For optimal algorithms (OA), their performance is usually evaluated through 23 benchmark functions (BF) [29,34]. In this paper, ISGA is compared to SGA and Particle Swarm Optimization (PSO) [34] via 23 benchmark functions.

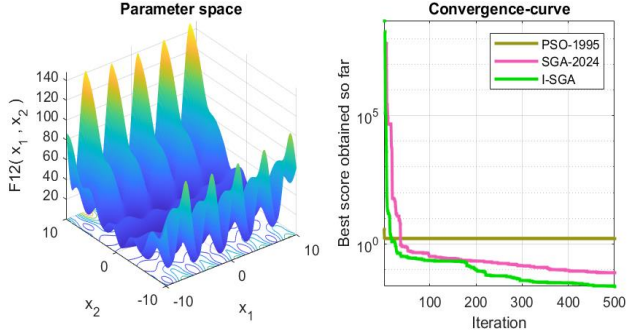

Fig. 2. Compare optimal performance on the 12th BF

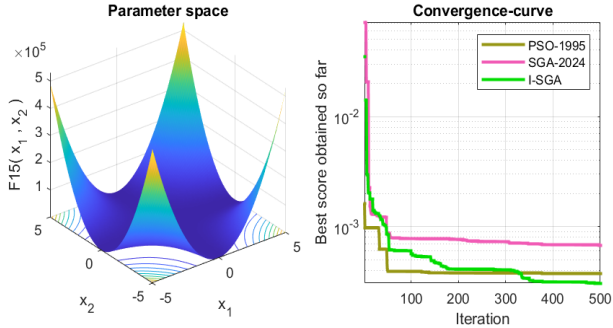

Fig. 3. Compare optimal performance on the 15th BF

The initial condition is set as follows: n=30; M=500; $f_{min}$=10; $f_{max}$=100. The results of the survey showed that ISGA is better than the SGA. Specifically, the performance comparison results on the 12$^{th}$ and 15$^{th}$ BF of the OA are illustrated in Figures 2 and 3, respectively. Survey results on 23 BF confirm that ISGA has better optimal searching than SGA. For example, for the 12$^{th}$ BF, the optimal values that PSO, SGA, and ISGA find are 1.669, 0.077012, and 0.022806, respectively.

## B. Performance Evaluation of ISGA-GMMEEF-AUKF

### Scenario 1: Random_Impulse_Q noise

In this scenario, the effect of Random_Impulse_Q noise on PS is considered, which is constructed with its full code in [26] and is described by:

$$\mathbf{q}_t \simeq N(0, \mathbf{I}_m) + Random\_impulse\_1 \quad (78)$$

$$\mathbf{r}_t \simeq N(0, \mathbf{R}) + Random\_impulse\_2 \quad (79)$$

where $N(*)$: Gaussian distribution. $\mathbf{I}_m$: The identification matrix has a size appropriate to the bus number of the PS; *Random_impulse_1* and *2*: outliers are added by adopting two functions *randn* and *randi* in Matlab.

Figure 4 displays the voltage phase (V-P) estimation error on IEEE 57-bus and Table II presents the voltage amplitude (V-A) estimation error for each IEEE system in this scenario.

TABLE II
VOLTAGE AMPLITUDE ESTIMATION ERROR FOR SCENARIO 1

| Algorithm \ IEEE | Voltage amplitude estimation error | | |
|---|---|---|---|
| | 14-bus | 30-bus | 57-bus |
| UKF | 0.008669 | 0.008709 | 0.008561 |
| MCC-UKF | 0.008415 | 0.008375 | 0.008398 |
| MEE-UKF | 0.008434 | 0.008434 | 0.008214 |
| MEEF-UKF | 0.008294 | 0.008267 | 0.008009 |
| GMMEEF-AUKF | 0.007507 | 0.007614 | 0.007230 |
| ISGA-GMMEEF-AUKF | **0.007396** | **0.007405** | **0.007115** |

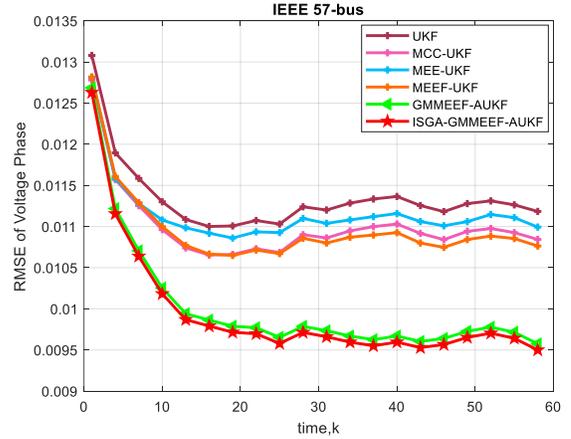

Fig. 4. RMSE of V-P on IEEE 57-bus for scenario 1.

It can be easily confirmed that the proposed algorithms GMMEEF-AUKF and ISGA-GMMEEF-AUKF have achieved superior performance compared to the existing algorithms. When confronted with Random_Impulse_Q noise, current algorithms such as MCC-UKF, MEE-UKF, and MEEF-UKF all demonstrate improved performance compared to traditional UKF. However, this difference is not so striking, similar to MCC-EKF versus EKF in [26]. It can be said that these algorithms do not meet the current requirements.

On the contrary, the proposed algorithms with the flexibility of the generalized Gaussian kernel and the adaptive update step of the noise covariance matrix have easily overcome the influence of Random_Impulse_Q noise. These achieved results once again confirm the contributions in the studies [9-15].

### Scenario 2: Bimodal Gaussian mixture noise and outliers

During the exploitation, the impact of the unexpected events on PS may be described as a bimodal Gaussian mixture noise and outliers [4,8,35]. It should be noted that the Gaussian distribution with small probability and a large variance such as $N(0.2,0.3)$ and $N(0,20)$ can be considered outliers (impulsive noises) [3]. Measurement and process noises are modeled by:

$$\mathbf{q}_t \simeq 0.4N(0.2, 10^{-4}) + 0.2N(0, 10^{-2}) + 0.4N(-0.2, 10^{-4}) \quad (80)$$

$$\mathbf{r}_t \simeq 0.4N(0.2, 0.3) + 0.2N(0, 20) + 0.4N(-0.2, 0.3) \quad (81)$$

Figure 5 displays the V-P estimation error on IEEE 30-bus. Figure 6 illustrates the V-A estimation error for each IEEE system in this scenario. When estimating the PS under the impact of bimodal Gaussian mixture noise and outliers, GMMEEF-AUKF and ISGA-GMMEEF-AUKF have shown their robustness compared to MEEF-UKF. MCC-UKF and

MEE-UKF achieved not high performance because of the poor flexibility of MCC and MEE when faced with the complicated distribution of noise. On IEEE 30-bus system, the efficiency of ISGA-GMMEEF-AUKF is higher than GMMEEF-AUKF, MEEF-UKF, ISGA-AUKF, and UKF are 17.5%, 30.9%, 61%, 68.4%, respectively.

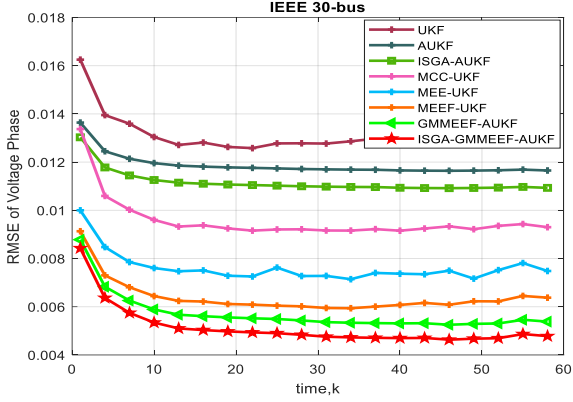

Fig. 5. RMSE of V-P on IEEE 30-bus for scenario 2.

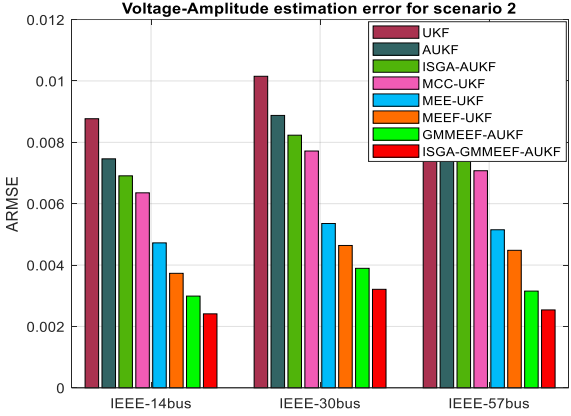

Fig. 6. ARMSE of V-A on IEEE-14,30,57bus for scenario 2.

*Scenario 3: Asymmetry Gaussian mixture noise and outliers*

Different from scenario 1, the impact of the unexpected events on PS may be described as an asymmetry Gaussian mixture noise and outliers [4,8,35]. Measurement and process noises are modeled as follows:

$$\mathbf{q}_t \simeq 0.4N(0.3,10^{-3}) + 0.2N(0,10^{-2}) + 0.4N(-0.1,10^{-4}) \quad (82)$$

$$\mathbf{r}_t \simeq 0.4N(0.3,0.2) + 0.2N(0,20) + 0.4N(-0.1,0.3) \quad (83)$$

Figure 7 displays the voltage amplitude (V-A) estimation error on IEEE 14-bus and Figure 8 illustrates the V-P estimation error for each IEEE system in this scenario.

When faced with asymmetry Gaussian mixture noise and outliers, GMMEEF-AUKF and ISGA-GMMEEF-AUKF still show their robustness. MCC-UKF, MEE-UKF, and MEEF-UKF with limitations that have been analyzed when estimated have the performance is lower than the proposed algorithms. The above three scenarios confirm the great influence on the performance of coefficients ($\alpha$, $\beta$, $\alpha_1$, $\alpha_2$, $\alpha_3$, $\beta_1$, $\beta_2$, $\beta_3$, $\theta$) in the algorithms based on learning criteria. Through ISGA easily finds the optimal coefficients fit for the PS model, and improves the estimated performance. On the IEEE 57-bus system, ISGA-GMMEEF-AUKF has 12.6% better efficiency than GMMEEF-AUKF and 27% than MEEF-UKF.

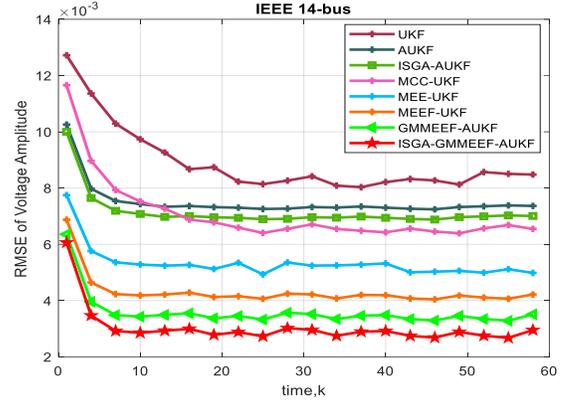

Fig. 7. RMSE of V-A on IEEE 14-bus for scenario 3.

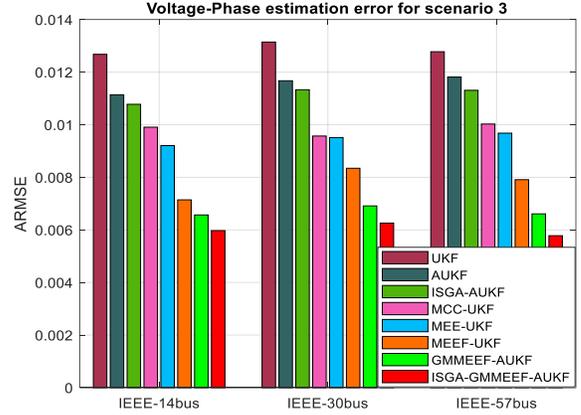

Fig. 8. ARMSE of V-P on IEEE-14,30,57bus for scenario 3.

Scenario 4: *Bad measurement data*

In scenario 4, similar to in [10,15], power measurement data, which includes real power and reactive power, is affected by outliers considered. Let's assume that at the 20$^{th}$ and 40$^{th}$ times, the data obtained on power measurements are increased by 15% and decreased by 15%, respectively. Besides, the electrical system is also affected by the same noise as in scenario 1. The results of the V-A state estimation and error on the IEEE 57-bus system are shown in Figures 9 and 10, respectively.

Based on the results obtained, it can be observed that the proposed algorithm still obtains the highest accuracy. In a complex scenario consisting of affected measurement data and under the influence of non-Gaussian noise and outliers, the two algorithms GMMEEF-AUKF and ISGA-GMMEEF-AUKF have achieved excellent performance. The reason is because their optimal criteria are built on mixture correntropy [14,15]. In contrast, the UKF, AUKF, and ISGA-AUKF algorithms are built on Gaussian assumptions, so their performance is greatly affected in this scenario. On IEEE 30-bus system, the efficiency of ISGA-GMMEEF-AUKF is higher than MEEF-UKF, MEE-UKF, MCC-UKF, ISGA-AUKF and UKF are 24%, 31%, 43%, 58%, 62%, respectively.

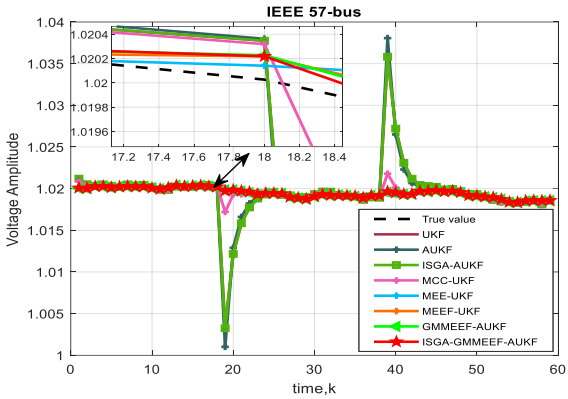

Fig. 9. Estimate V-A on IEEE 57-bus for scenario 4.

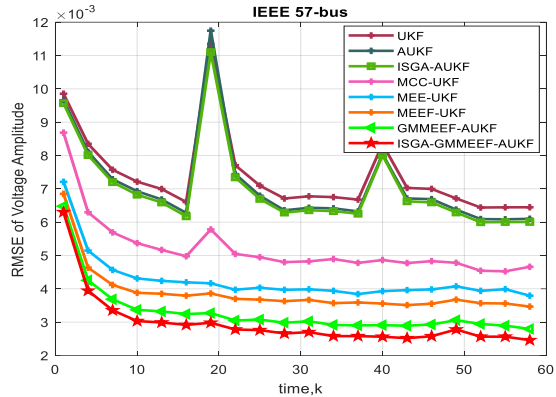

Fig. 10. RMSE V-A on IEEE 57-bus for scenario 4.

### C. Compare Computing Time

To compare the estimated time of the algorithms, test conditions are set the same as in scenario 1. The single-step running time of the estimation algorithms is given in Table III.

It can be commented that the ISGA-GMMEEF-AUKF algorithm needs time to calculate each step more. However, this is completely acceptable compared to the accuracy and flexibility that ISGA-GMMEEF-AUKF has achieved.

TABLE III
COMPARE RUNNING TIMES

| IEEE Algorithm | Single-step running time (s) | | |
|---|---|---|---|
| | 14-bus | 30-bus | 57-bus |
| UKF | 0.0053 | 0.0371 | 0.1866 |
| ISGA-AUKF | 0.0072 | 0.0394 | 0.1903 |
| MEEF-UKF | 0.0214 | 0.0935 | 0.3679 |
| GMMEEF-AUKF | 0.0235 | 0.1119 | 0.3912 |
| ISGA-GMMEEF-AUKF | 0.0252 | 0.1225 | 0.4084 |

## VI. CONCLUSION

In this paper, the algorithms GMMEEF-AUKF and ISGA-GMMEEF-AUKF have been proposed for the DSE of PS. Through survey results of IEEE systems on different scenarios, the excellent performance has been confirmed. The modified Sage-Husa estimator technique has been used to update online the noise covariance matrices, which satisfy the requirements of DSE. The main pivotal is the GMMEEF criterion which the GMMC has put into the GMEE cost function will help GMMEEF automatically locate the vertex of the error PDF and fix it at the origin more flexible than MEEF. On the other hand, the ISGA has been designed to eliminate the difficulty of choosing the optimal value of the kernel shape coefficients in the learning criteria, the selection of sigma points in UT, and the update coefficient of the noise covariance matrices. In the future, a distributed estimation version, which uses multi-sensor networks and consensus algorithms, will be developed to meet the growing scale of the power system.